\documentclass[twocolumn,showpacs,preprintnumbers,amsmath,amssymb]{revtex4}
\usepackage{graphicx}
\usepackage{dcolumn}
\usepackage{amsmath}
\usepackage{epsfig}

\begin{document}
\title{Inhibiting decoherence via ancilla processes}
\author{Guihua Zeng}%
\email{guihuazeng@hotmail.com}%
\author{Christoph H. Keitel}%
\email{keitel@uni-freiburg.de}%
\affiliation{Theoretische Quantendynamik, Fakult\"at f\"ur Physik,
Universit\"at Freiburg, Hermann-Herder-Stra{\ss}e 3, D-79104 Freiburg, Germany}

\date{\today}

\begin{abstract}

General conditions are derived for preventing the decoherence of a single two-state quantum system (qubit) in a thermal
bath. The employed auxiliary systems required for this purpose are merely assumed to be weak for the general condition
while various examples such as extra qubits and extra classical fields are studied  for applications in quantum
information processing. The general condition is confirmed with well known approaches towards inhibiting decoherence.
A novel approach for decoherence-free quantum memories and quantum operations is presented by placing the qubit into
the center of a sphere with extra qubits on its surface.

\end{abstract}
\pacs{03.65.Yz, 03.67.-a, 03.65.Ta}%
\maketitle

\section{Introduction}

The physical quantity {\it coherence} is a measure of the coupling between quantum mechanical states and represents an
elementary phenomenon in the microscopic and macroscopic world \cite{divincenzo95,myatt00}. It is associated with the
well defined superposition of quantum mechanical states and thus with fundamental properties of quantum mechanics. The
coherence of a state is generally defined for a closed system, i.e. without the inclusion of external influences.
However, in virtually all cases an environment will act on the system of interest and modify, i.e. generally reduce
its coherence. The infinite-mode state of the environment will become entangled with the superposition state of the
system and, subsequently, will deteriorate the coherence of the system \cite{shorzurek95}, a process called
decoherence.

Decoherence has been studied as an essential characteristic of quantum superposition systems in many fundamental works
\cite{unruh95,zurek81,haba98,gisin84,agarwal99} while in fact it plays a crucial role in virtually all applications in
modern quantum optics, quantum information and quantum mechanics. Usually, decoherence is viewed as passive, leading
unintendedly to the leakage of information from the system. In particular couplings between quantum states and
interference effects are very vulnerable to actions of the deteriorating influence of the environment. For example in
quantum information processing errors are introduced by the environment to a serious degree. Equally atomic coherence
effects such as lasing without inversion and spontaneous emission control \cite{sec} and also laser cooling
\cite{lascool} are limited by the influence of the environment. These examples are just a few of a long list and are
only meant to show the importance of finding means of controlling decoherence.

In order to prevent decoherence two representative approaches have been proposed in quantum information theory. One of
them is the quantum error correction approach \cite{shor95,steane96,plenio97}, which is employed to detect and/or
correct the errors induced by the environment. The second one is the quantum error avoiding approach \cite{zanardi97,
duan97, lidar98}, which is employed to avoid the errors induced by decoherence or any kinds of noise. This approach
constructs a decoupled subspace which is called decoherence-free subspace (DFS) for the system with the consequence
that errors can be avoided if the system is placed in the DFS. Physically, the DFS is a subspace of the system's
Hilbert space, where the evolution is decoupled from the environment. Several approaches have been proposed to
construct such a DFS \cite{alber01, viola97,poyatos96}, employing various means to decouple the system from the
environment. While all approaches are aimed to establish a  DFS in the total space of the Liouvillian evolution of the
system state, there should be a united and general condition for constructing the DFS and to inhibit decoherence.

In this article a general condition is derived to construct a DFS for a two state quantum system in a thermal bath
with the aid of a weak assistant system. Emphasis is also placed on relating those conditions to realistic physical
situations, especially on applying them to present-day problems in quantum information processing. The assistant
systems considered to inhibit decoherence may be extra ensembles of qubits, extra classical fields or combinations of
them. The additional systems are shown in particular to allow for decoherence-free quantum memories or quantum
operations. The general conditions are in agreement with previously proven means of inhibiting decoherence while
further examples are presented as up-to-date unexplored means. In particular the decoherence of qubits may be
eliminated by placing them into the center  of a sphere with additional qubits surrounding it on the surface.

In Sect. \ref{dyn}, our investigations begin with the description of general aspects of decoherence of a two state
system $\cal S$ in a thermal bath $\mathcal B$. DFS are introduced in Sect. \ref{workeq} followed by the derivation
of a general condition for the construction of a DFS with a weak but otherwise general auxiliary system.
A particular auxiliary system built from extra qubits and extra fields is employed in Sect. \ref{app} and in separate
subsections applied to decoherence-free quantum memories and operations. The article concludes in Sect. \ref{con}.

\section{Dynamics with decoherence}
\label{dyn}

Before showing  how to prevent decoherence in the following sections we first investigate its origin and the dynamics
it imposes on a quantum system. A fully coherent superposition of quantum systems $|\Psi\rangle=\sum_ic_i|i\rangle$ is
generally considered at the onset of the interaction with a bath which induces the decoherence. With time this
interaction  transforms the wave function into a state which need be described by a density operator rather than a
wave function, i.e.
\begin{eqnarray}
\rho=\rho_{m}+\sum_{i\neq j}c_ic_j|i\rangle\langle j|,
\end{eqnarray}
where $\rho_{m}=\sum_i|c_i|^2|i\rangle\langle j|$ describes a  classical mixture, i.e. a density operator
with purely diagonal terms. When the second term in the above expression decays to zero the coherent superposition
state becomes a fully mixed state, indicating that  the coherence has completely decayed.
The decoherence is associated with the decay of the off-diagonal terms in the density matrix, which in system-reservoir
interactions may often be described via the simple law \cite{braun01}
\begin{eqnarray}
\frac{\partial \langle i| \rho |j\rangle}  {\partial t} =  \frac{\partial \rho_{ij}}{\partial t}  =-f(t)\rho_{ij},
\end{eqnarray}
where $f(t)=D(t)+i\Phi(t)$, and $D(t)$ is the decoherence rate.

For simplicity, we here investigate decoherence of a two-state system $\cal S$ in a thermal bath $\cal B$. The bath is
modeled by oscillators with infinite degrees of freedom being described by annihilation (and creation) operators
$b_{\lambda}$ (and $b^\dag_{\lambda}$) and density distributed frequencies $\omega_\lambda$. The two state system is
modeled by a superposition of an excited state $|2\rangle$ and a ground state $|1\rangle$, i.e. $
|\Psi\rangle=c_1|1\rangle+c_2|2\rangle$, with a resonant transition frequency $\omega$ (basic qubit). In such  a
situation the interaction Hamiltonian of the total system has the following form in the interaction picture within the
rotating wave approximation and the dipole approximation,
\begin{equation}\label{eq1:MECH-Hsb}
\mathcal H_{sb}=\Gamma^\dag_a\sigma^-+\Gamma^\dag_p\sigma_z +H.c.,
\end{equation}
where $\Gamma^\dag_a$ and $\Gamma^\dag_p$ are the bath operators and may be expressed as
sums over all modes of the baths (parametrised by $\lambda$)
\begin{equation}\label{eq1:MECH-Gamma}
\Gamma^\dag_a=\sum_\lambda g_{a\lambda} b^\dag_\lambda, \,\,\,\,\,%
\Gamma^\dag_p=\sum_\lambda g_{p\lambda} b^\dag_\lambda,
\end{equation}
and
\begin{equation}\label{eq1:MECH-Gamma*}
\Gamma_a=(\Gamma^\dag_a)^*,\,\,\,\,\,%
\Gamma_p=(\Gamma^\dag_p)^*.
\end{equation}
$g_{\alpha\lambda}$ and $g^*_{\alpha\lambda} (\alpha\in\{a,p\})$ are the corresponding coupling coefficients between
the system and the bath. $\sigma^+,\sigma^-$ and $\sigma_z$ are the atomic operators defined via
\begin{equation}\label{eq1:MECH-sigma}
\sigma^+=|2\rangle\langle 1|, \,\,\,\,\, \sigma_z=|2\rangle\langle 2|-|1\rangle\langle 1|,
\end{equation}
and
\begin{equation}\label{eq1:MECH-sigma*}
\sigma^-=(\sigma^+)^*=|1\rangle\langle 2|.
\end{equation}
Employing the Markov approximation, the corresponding master equation of the two-state system in the interaction
picture may be written as \cite{walls95}
\begin{eqnarray}\label{eq1:MECH-master}
\frac{\partial \rho}{\partial t} %
&=& \mathcal L_0 \rho\nonumber\\
&=& \frac{\gamma_\shortparallel}{2}(N+1)\left([\sigma^-\rho, \sigma^+]
+[\sigma^-,\rho\sigma^+]\right)\nonumber\\ %
&+& \frac{\gamma_\shortparallel}{2}N\left([\sigma^+\rho, \sigma^-]
+[\sigma^+,\rho\sigma^-]\right)\nonumber\\ %
&+&\frac{\gamma_p}{4}\left([\sigma_z\rho, \sigma_z] +[\sigma_z,
\rho\sigma_z]\right),
\end{eqnarray}
where $\mathcal L_0$ is the Liouvillian operator, $\gamma_\shortparallel$ and $\gamma_p$ describe the amplitude and
phase decay rates described by the following forms,
\begin{equation}\label{eq1:Mech-gamma}
\gamma_\alpha=2\pi \sum_\lambda g^2_{\alpha\lambda} \delta (\omega_\lambda-\omega).
\end{equation}
Under the condition of a thermal equilibrium, $N$ has the following form
\begin{equation}\label{eq1:Mech-N}
N=\left [e^{\hbar \omega/k_B T}-1 \right ],
\end{equation}
with $T$ being the temperature of the environment, and $k_B$ the Boltzmann constant.

Formulating  Eq. (\ref{eq1:MECH-master}) in terms of the matrix elements
yields
\begin{equation}\label{eq1:MECH-rho11}
\frac{\partial \rho_{11}}{\partial t}=-\gamma\rho_{11}+\theta,
\end{equation}
and
\begin{equation}\label{eq1:MECH-rho12}
\frac{\partial \rho_{12}}{\partial t}=-D\rho_{12},
\end{equation}
where
\begin{equation}\label{eq2:MECH-gamma}
\gamma=\gamma_\shortparallel(2N+1),
\end{equation}
and
\begin{equation}\label{eq1:MECH-theta}
\theta=\gamma_\shortparallel (N+1).
\end{equation}
Consequently, the off-diagonal term $\rho_{12}$ evolves with the following decoherence rate
\begin{equation}\label{eq1:MECH-D}
D=\frac{\gamma_\shortparallel}{2}(2N+1)+\gamma_p.
\end{equation}
Furthermore Eq.(\ref{eq1:MECH-rho12}) can be solved easily to give
\begin{equation}\label{eq2:MECH-rho12}
\rho_{12}=e^{-Dt}\rho^0_{12},
\end{equation}
where $\rho^0_{12}$ indicates the initial value of the coherence at time $t=0$.

We note that the decoherence rate $D$ depends on the coupling coefficients $g_{\alpha\lambda}$. In addition,
$g_{\alpha\lambda}\rightarrow 0$ leads to $D\rightarrow 0$, and consequently $\rho_{12}=\rho^0_{12}$, i.e. the
coherence of the system is preserved. Accordingly, the decoherence is generated by the coupling between the system and
the environment. This kind of coupling establishes an  entanglement between the system and environment, however
induces decoherence of the system. In order to prevent decoherence, the system must be decoupled from the environment.
As we see later this can be implemented by a further interaction which compensates for the interaction with the bath.

Eq.(\ref{eq1:MECH-D}) also shows that the decoherence rate $D$ is associated with the temperature of the environment.
In the low temperature case the decoherence parameter $D$ satisfies the condition
\begin{equation}\label{eq2:MECH-D}
D_1=N\gamma_\shortparallel+\gamma_p.
\end{equation}
In the high temperature case the decoherence parameter $D$ becomes
\begin{equation}\label{eq3:MECH-D}
D_2=\frac{\gamma_\shortparallel}{2}+\gamma_p=\gamma_\perp.
\end{equation}
We note that the decoherence is only dependent on the transverse coupling coefficient $\gamma_\perp$. In a general
situation, the decoherence rate decreases exponentially with the environmental temperature.

\section{Inhibition of Decoherence}
\label{workeq}

In this section we present how to obtain decoherent free subspaces in principle and then derive working
equations which display the precise conditions on the auxiliary system to be fulfilled to inhibit the decoherence.

\subsection{Decoherence-free subspaces}
\label{DFS}

In previous section we have shown decoherence to arise from the entanglement between the system and the environment.
In additional $D\rightarrow 0$ was associated with coherence being preserved. In what follows we will derive
conditions to obtain $D\rightarrow 0$. This will have to be with the aid of an addition interaction because
Eq.(\ref{eq1:MECH-D}) excludes any other possibility.

In the complete manifold of quantum states $\mathcal H_s$, a submanifold $\mathcal M_s$ of fixed points (stationary
states) of the Liouvillian evolution was shown to exist. Consequently  the coherence is preserved in $\mathcal M_s$,
or in other words the dynamics in  $\mathcal M_s$ is a fortiori unitary \cite{zanardi97}. Since  decoherence is absent
in such a submanifold $\mathcal M_s$, it has been called  decoherence free subspace (DFS) \cite{lidar98}. Although DFS
have been proven to exist, a general condition for the existence of DFS and the corresponding practical implications
remain a  key problem in practice. Obviously, in Eqs. (\ref{eq1:MECH-rho11}, \ref{eq1:MECH-rho12}) the stationary
state are,
\begin{equation}\label{eq1:DFS-rhoij}
\rho_{11}=\frac{\theta}{\gamma}, \,\,\,\,\, \rho_{12}=0,
\end{equation}
which can not construct a DFS since the off-diagonal terms are vanishing. Thus, to prevent decoherence or build a DFS,
the construction of the Hamiltonian of the total system ($\cal S+\cal B$) must be modified by an additional physical
process or an assistant system $\cal A$.

For the sake of generality we consider here a general assistant system $\cal A$, which may be associated with any
possible physical process. For example, assistant systems $\cal A$ which may be employed to modify the environment,
drive the total system by an electric field or change the distribution of the system of qubits, are all possible in the
proposed model. From the point of view of practical application, a weak additional system is more advantageous than a
strong additional system, because a weak additional interaction can be more easily controlled and implemented
experimentally, often also with a reduced cost. We here employ a weak additional system $\cal A$ to control the
decoherence. In this situation the contribution of the assistant system can be modeled as a perturbation $\mathcal
H_p$ with respect to the  system (${\cal S}$) to be placed in a DFS.

To build a DFS we start with the master equation with a perturbation $\mathcal H_p$. Then the reduced density of the
system $\cal S$ in the interaction picture $\rho^s$ can be written as
\begin{equation}\label{eq1:DFS-master}
\frac{\partial \rho^s}{\partial t} = \mathcal L_0 \rho^s -
\frac{i}{\hbar}[\tilde{\mathcal H}_I, \rho^s]=\mathcal L \rho^s,
\end{equation}
where $\mathcal L=\mathcal L_0-i/\hbar[\tilde {H}_I, \cdot]$, $\mathcal L_0$ is
defined in Eq.(\ref{eq1:MECH-master}), and
\begin{equation}\label{eq1:DFS-Hp}
\tilde{\mathcal H}_I=e^{\frac{i}{\hbar}\mathcal H_st} \mathcal H_p e^{-\frac{i}{\hbar}\mathcal H_st},
\end{equation}
with $\mathcal H_s$  being the free Hamiltonian of the system $\cal S$. According to the definition of the DFS, the
sufficient and necessary condition for building a DFS is
\begin{equation}\label{eq1:DFS-condition}
\frac{\partial \rho^s}{\partial t} = \mathcal L\rho^s=0,
\end{equation}
or,
\begin{equation}\label{eq2:DFS-condition}
[\tilde{H}_I, \rho^s]=-i\hbar\mathcal L_0 \rho^s.
\end{equation}
Above expression shows the existence of stationary states of the system  when the assistant system is chosen
appropriately. Therefore, the modified Liouvillian operator $\mathcal L$ constructs a DFS at the condition of Eq.
(\ref{eq2:DFS-condition}).

Obviously the constructed DFS depends on the parameters of the additional system. Choosing an appropriate assistant
system the coherence can be preserved in the constructed DFS. In this situation the decoherence time is
\begin{equation}
\tau_d=\sum_i\tau_i,
\end{equation}
where the $i^{th}$ order decoherence time can be denoted as \cite{lidar98},
\begin{equation}\label{eq1:DFS-ntime}
\tau_i=\left\{Tr\left(\rho(t_0)(\mathcal L)^i \rho(t_0)\right)\right\}^{-1/i},
\end{equation}
with $t_0$ being a  time in  the stationary regime of the dynamics. From Eqs. (\ref{eq1:DFS-condition}) and
(\ref{eq1:DFS-ntime}) one can easily infer that $\tau_1\rightarrow \infty$, while $1/\tau_i\neq 0$  for $i\geq 2$.
This means that coherence can be preserved in the DFS only to the first order. With the decoherence time $\tau_d$
being larger than the first order decoherence time  $\tau_1$ however, it is sufficient to reach a DFS, i.e.
$\tau_1\rightarrow \infty$, to avoid any decoherence, i.e. reach $\tau\rightarrow \infty$.

\subsection{General conditions for preserving coherence}
\label{gencon}

In the previous subsection \ref{DFS} we showed that it is possible in principle to construct a DFS by
introducing an appropriate additional interacting system. In this subsection we consider the features
of the additional system, i.e., the construction of the additional Hamiltonian $\mathcal H_p$ in some more detail.
Eq. (\ref{eq1:DFS-master}) may be written in terms of the matrix elements of the density operator of our two-state
system $\rho^s$
\begin{equation}\label{eq1:cond-rho12}
\frac{\partial \rho^s_{12}}{\partial t}=-D\rho^s_{12}-\frac{i}{\hbar}\Delta H
\rho^s_{12}-\frac{i}{\hbar}H_{12}(1-2\rho^s_{11}).
\end{equation}
and
\begin{equation}\label{eq1:cond-rho11}
\frac{\partial \rho^s_{11}}{\partial t}=-\gamma\rho^s_{11}+\theta
-\frac{i}{\hbar}\left(H_{12}\rho^s_{21}-H_{21}\rho^s_{12}\right),
\end{equation}
where $H_{ij}=\langle i|\tilde{\mathcal H}_I|j\rangle,$ with $ \{i,j\} \in \{1, 2\}$ and
$\Delta H=H_{11}-H_{22}$. Combining Eq.
(\ref{eq1:DFS-condition}) and Eq. (\ref{eq1:cond-rho12}) gives
\begin{equation}\label{eq1:cond-H12}
H_{12}=\frac{\rho^s_{12}}{1-2\rho^{s}_{11}}\left (i\hbar D-\Delta H\right )
\end{equation}
with $H_{21}=H^*_{12}$.

Inserting Eqs.(\ref{eq1:cond-H12}) and the corresponding one for its complex conjugate
into Eq.(\ref{eq1:cond-rho11}), one obtains
\begin{equation}\label{eq2:cond-rho11}
\frac{\partial \rho^s_{11}}{\partial t}=-\gamma\rho^s_{11}+\theta
-\frac{|\rho^s_{12}|^2}{1-2\rho^s_{11}}\left(\kappa- 2D\right),
\end{equation}
where
\begin{equation}\label{eq1:cond-kappa}
\kappa=\frac{i}{\hbar}(\Delta H^*-\Delta H).
\end{equation}
Since $\rho^s_{11}, |\rho^s_{12}|^2$ and the parameters $D, \gamma, \theta$ are real numbers, Eq. (\ref{eq2:cond-rho11})
demands the parameter $\kappa$ to be a real number as well. Combining Eqs. (\ref{eq1:DFS-condition}) and
(\ref{eq2:cond-rho11}) yields,
\begin{equation}\label{eq3:cond-rho11}
2\gamma(\rho^s_{11})^2-(\gamma+2\theta)\rho^s_{11}+\theta- (\kappa-2D)|\rho^s_{12}|^2=0.
\end{equation}
This is a quadratic equation for $\rho^s_{11}$ with 0, 1 or 2 solutions.
We requite only one solution for $\rho^s_{11}$ and thus impose
\begin{equation}\label{eq1:cond-Delta}
(\gamma+2\theta)^2-8\gamma[\theta-(k-2D)|\rho^s_{12}|^2]= 0.
\end{equation}
Then Eqs.(\ref{eq1:cond-H12}), (\ref{eq3:cond-rho11}) and
(\ref{eq1:cond-Delta}) yield
\begin{equation}\label{eq2:cond-H12}
H_{12}=\sqrt{\frac{\gamma}{2(2D-\kappa)}}\left |i\hbar D-\Delta H \right
|e^{i\varphi},
\end{equation}
where $\varphi$ is the phase of $H_{12}$.

In order to inhibit decoherence, we thus need impose conditions only on the diagonal elements  $H_{21}$ and $H_{12}$
of the additional Hamilton operator  $\tilde{\mathcal H}_I$. We note that the additional Hamiltonian is arbitrary in
the sense that its diagonal elements $H_{11}$ and $H_{22}$ as well as the phase $\varphi$ of $H_{12}$ are not
determined. We will thus not restrict them; they do not play any role in the further investigations in the following
section on applications.

Thus Eq. (\ref{eq2:cond-H12}) represents a sufficient general condition for the preservation of the coherence of a
two-state system in a thermal reservoir. The assistant system $\cal A$ with corresponding Hamiltonian $\mathcal H_p$
has not been associated so far with a concrete physical process or system and in principle of course one may choose
not to do so. However, $\cal A$ may be a concrete realistic physical system, e.g. for modulating the environment. It
may be associated with an electromagnetic field driving the system $\cal S$, an ensemble of several other qubits or any
combination of these systems. In addition to fulfill Eq. (\ref{eq2:cond-H12}) we require $\mathcal H_p$ to be a
perturbation only which shall be useful in realizing it in practice. To some degree we may also restrict ourselves to
the additional system $\cal A$ to prevent the decoherence, such as e.g. an external field $\cal F$ and an extra system
of qubits $\cal Q$. Then we need choose at least the number of the additional qubits appropriately to fulfill Eq.
(\ref{eq2:cond-H12}) and in addition the parameters of the additional fields, such as the frequency and the intensity
of the laser field to control the system. Thus all we need is that the perturbation Hamiltonian of the additional
system $\cal A$ and the information system ${\cal S}$ satisfy jointly Eq. (\ref{eq2:cond-H12}). Then the decoherence
can be controlled. i.e. the coherence can be preserved. The special model involving coherent fields and extra qubits
will be discussed in the next section in detail.

In the following we consider the special situation of an ancilla
system with $\Delta H=0$. Then Eq.(\ref{eq2:cond-H12}) can be simplified to give
\begin{equation}\label{eq3:cond-H12}
H_{12}=\frac{\hbar}{2}\sqrt{D\gamma}e^{i\varphi}.
\end{equation}
Thus the Hamiltonian $\tilde{\mathcal H}$ in the interaction picture can be
written as,
\begin{equation}\label{eq4:cond-H'}
\tilde{H}_I=\left(
\begin{array}{cc}
H_{11} & \frac{\hbar}{2}\sqrt{D\gamma}e^{i\varphi}\\
\frac{\hbar}{2}\sqrt{D\gamma}e^{-i\varphi} & H_{22}
\end{array}
\right).
\end{equation}
In a high temperature environment we have $ k_BT \gg \hbar \omega_0$, which means
$N\rightarrow 0$. Then Eqs. (\ref{eq2:MECH-gamma}), (\ref{eq1:MECH-D}) and
(\ref{eq3:cond-H12}) yield
\begin{equation}
H_{12}= \frac{\hbar}{2}\sqrt{\gamma_\shortparallel\gamma_\perp} e^{i\varphi},
\end{equation}
which means that $H_{12}$ involves a coupling of the parallel and transverse decays. This reminds us
of interference effects in the spontaneous emission of a doublet of quantum states which are closely spaced
with regard to their line width so that they couple to the same vacuum modes \cite{sec}. In the low temperature case we
obtain $N\gg 1$. Then Eqs. (\ref{eq2:MECH-gamma})
(\ref{eq1:MECH-D}) and (\ref{eq3:cond-H12}) lead to
\begin{equation}
H_{12}= \frac{\hbar}{2}\sqrt{2N\gamma_\shortparallel(N\gamma_\shortparallel
+\gamma_\perp)} e^{i\varphi}.
\end{equation}

In the following section we will connect the more formal conditions presented so far for the ancilla interaction
Hamiltonians with concrete realistic physical situations where decoherence can be inhibited.

\section{Applications in quantum information processing}
\label{app}

The investigations on decoherence presented so far may be applied to the understanding of the basic conception
of quantum mechanics or to the large number of novel effects in modern quantum optics and quantum information
where decoherence is a serious problem. We here concentrate on applications in quantum information processing.
Currently, two key problems in this field associated with decoherence appear to be information storage and
operations on information. In order to keep the qubit unaltered in a quantum memory or to transform it in a
quantum operation without information loss, the decoherence time $\tau_d (\approx\tau_1)$ should be longer than
the time $\tau_c$ of the qubits interacting with the environment, i.e, $\tau_d\geq \tau_c$.
It has been shown so far that this relationship can be implemented by two major approaches, which are the quantum
error correction approach and the quantum error avoiding technologies as mentioned in the introduction.

Physically, the quantum error correction approach controls decoherence employing the entanglement of involved quantum
states. In particular, the information qubits are entangled employing especially prepared qubits such as $|0\rangle$
or $|1\rangle$ in addition to quantum operations with the effect that the errors arrising due to decoherence are
transfered to the additional qubits. The decoherence in the quantum error avoiding approach is controlled by constructing
a DFS. Usual ways for the generation of a DFS are the employment of an external field, the change of the arrangement of
qubits, or the engineering of the environment as mentioned in the introduction. All these approaches can be
described by a general model, i.e., an assistant system $\mathcal A$. In the above section, we have obtained a
general condition (see Eq. (\ref{eq2:cond-H12})) for building a DFS. In this section, we discuss how to apply the
resulting condition in quantum information and quantum computation. We will rederive known results with our
approach and put forward new mechanisms based on our formalism.

\subsection{The employed ansatz for the ancilla system}

Most applications for decoherence inhibition, including the ones presented in the two following subsections, involve
extra qubits, extra fields or extra quantum operations. We thus consider in this subsection the restricted though
still rather general model, in which the additional system is assumed to consist of an ensemble of additional qubits
denoted by $\mathcal Q$, a system of fields denoted by $\mathcal F$ and diverse quantum operations $\mathcal O$, such
as quantum gates $\mathcal G$. Then the perturbation $\mathcal H_p$ can be written as
\begin{equation}\label{eq1:MOD-H'}
\mathcal H_p=\mathcal H_{qs}+\mathcal H_{fs}+\epsilon \mathcal H_g,
\end{equation}
where $\mathcal H_{qs},\mathcal H_{fs}$ and $\mathcal H_g$ denote respectively the perturbation Hamiltonian induced by
the system $\cal Q$, the fields $\mathcal F$ and quantum gate operations $\mathcal G$; the parameter $\epsilon$
may take the value $0$ or $1$, so that we may also exclude gate operators as done in the first subsection to follow.
$\mathcal H_{qs}$ can be expressed as follows in the interaction picture
\begin{equation}\label{eq1:MOD-Hqs}
\mathcal H^I_{qs}=\sum^K_{k=1}\left (g_{sk}\sigma^+\sigma^-_ke^{i\delta_k} +
g^*_{sk}\sigma^-\sigma^+_ke^{-i\delta_k}\right ),
\end{equation}
where $\sigma^{+(-)}_k$ are the atomic operators of the $k^{th} (k\in \{1,2, \cdots, K\})$ qubit in the system
$\mathcal Q$, $\delta_k=\omega_k-\omega$, $\omega_k$ is the resonant transition frequency of the $k^{th}$ qubit in
$\mathcal Q$, $g_{sk}$ are the coupling coefficients between the system $\cal S$ and the k$^{th}$ qubits in $\mathcal
Q$ and $K$ is the number of qubits in the system $\mathcal Q$. In matrix form, Eq. (\ref{eq1:MOD-Hqs}) can be
rewritten as
\begin{equation}\label{eq2:MOD-Hqs}
\mathcal H^I_{qs}=\sum^K_{k=1}\left (
\begin{array}{cc}
g_{sk}e^{i\delta_k}  & 0\\
0 & g^*_{sk}e^{-i\delta_k}
\end{array}
\right).
\end{equation}
For convenience, we suppose the total external field $\mathcal F$ to be classical, so that in the interaction picture
$H_{fs}$ may be written as
\begin{equation}\label{eq1:MOD-Hfs}
\mathcal H^I_{fs}=\sum^L_{l=1}\left (\Omega_l \sigma^+ e^{i\delta_l}+\Omega_l^* \sigma^-e^{-i\delta_l}\right ),
\end{equation}
where $\Omega_l$ denote the Rabi frequencies of the $l^{th} (l\in \{1, 2, \cdots, L\})$ field in $\cal F$,
$\delta_l=\omega_l-\omega$, $\omega_l$ is the frequency of the $l^{th}$ external field in $\cal F$ and $L$ denotes the
number of the external fields.

Combining Eqs. (\ref{eq1:DFS-Hp}), (\ref{eq1:MOD-H'}), (\ref{eq2:MOD-Hqs}) and
(\ref{eq1:MOD-Hfs}) yields
\begin{equation}\label{eq2:MOD-H'}
\left(
\begin{array}{cc}
H_{11}-\sum_k\tilde{g}_{sk} & H_{12}-\sum_l\tilde{\Omega}_l \\
H_{21}-\sum_l\tilde{\Omega}^*_l & H_{22}-\sum_k\tilde{g}^*_{sk}
\end{array}
\right )=\epsilon \mathcal H^I_g,
\end{equation}
where $\tilde{g}_{sk}=g_{sk} e^{i\delta_k}$ and $\tilde{\Omega}_l=\omega_l e^{i\delta_l}$. In the resonant case, i.e.,
$\delta_k=0$ and $\delta_l=0$, Eq. (\ref{eq2:MOD-H'}) can be rewritten as
\begin{equation}\label{eq3:MOD-H'}
\left(
\begin{array}{cc}
H_{11}-\sum_kg_{sk} & H_{12}-\sum_l\Omega_l  \\
H_{21}-\sum_l\Omega_l^*  & H_{22}-\sum_kg^*_{sk}
\end{array}
\right )=\epsilon \mathcal H^I_g.
\end{equation}
Thus Eqs.(\ref{eq2:MOD-H'}) and (\ref{eq3:MOD-H'}) represent respectively a relationship in the non-resonant and the
resonant cases among $H_{qs}, H_{fs}$ and $H_g$, i.e., the relationship among the additional qubits $\cal Q$, the
additional fields $\mathcal F$ and the quantum operations $\mathcal O$. By means of this relationship one can
determine how to choose the additional system of qubits  $\mathcal Q$ and the driving fields $\mathcal F$ for the
decoherence preventation.

\subsection{Decoherence-free quantum memories}
\label{mem}

We begin by considering the problem of storing qubits for long times in a quantum memory in an unaltered state.
The quantum memory can be viewed as an environment to the system $\cal S$ to be stored. As a consequence,
the quantum memory will decohere the qubits and induce the information carried by the qubits to be lost.
In the following we will show how to preserve the coherence of the qubits according to our proposed approach,
i.e. by introducing an appropriate new interaction.

For a quantum memory, there is no quantum operation, i.e., $\epsilon=0$. Then Eq. (\ref{eq2:MOD-H'}) leads to
\begin{equation}\label{eq1:Memory-Hii}
H_{11}-\sum^K_{k=1}\tilde{g}_{sk}=0, \,\,\,\,\,\, H_{22}=H^*_{11},
\end{equation}
and
\begin{equation}\label{eq1:Memory-Hij}
H_{12}-\sum^L_{l=1}\tilde\Omega_l =0, \,\,\,\,\,\, H_{21}=H^*_{12}.
\end{equation}
Eqs.(\ref{eq1:Memory-Hii}) yield
\begin{equation}\label{eq1:Memory-DH}
\Delta H=2i\sum_k^K\Im(\tilde g_{sk}),
\end{equation}
and
\begin{equation}\label{eq1:Memory-kappa}
\kappa=\frac{4}{\hbar}\sum_k^K\Im(\tilde g_{sk}),
\end{equation}
where $\Im (\tilde g_{sk})=|\tilde g_{sk}-\tilde g^*_{sk}|$. Inserting Eqs. (\ref{eq1:Memory-Hij}),
(\ref{eq1:Memory-DH}) and (\ref{eq1:Memory-kappa}) into Eq. (\ref{eq2:cond-H12}), we obtain
\begin{equation}\label{eq1:Memory-condition}
\left|\sum^L_{l}\tilde\Omega_l\right|=\sqrt{\frac{\hbar^2\gamma \mu}{8}},
\end{equation}
where $\mu=2 D-\kappa$. By choosing an appropriate assistant system $\cal A$
containing of an additional system of qubits $\mathcal Q$ and/or a system of external fields  $\mathcal F$,
Eq.(\ref{eq1:Memory-condition}) shows that the coherence of the
qubit stored in quantum memory can be preserved by realistic means. In this situation, the qubit can be stored
in principle in the quantum memory for any requested time because the equation is equivalent to the first-order decoherence
time satisfying $\tau_1\rightarrow \infty$.

In what follows we consider three special situations. At first we would like to show that our approach is able to
reproduce well established results. Therefore we analyze the situation $K=1$ and $L=1$ in Eq. (\ref{eq1:Memory-condition}) ,
which turn out to be associated with the well-known particle-pairing approach \cite{duan97} for coherence preservation.
In this situation,
Eq.(\ref{eq1:Memory-condition}) gives,
\begin{equation}\label{eq1:Memory-driving}
|\tilde\Omega|=\sqrt{\frac{\hbar\gamma}{4}|\hbar D-2\Im (\tilde g)|},
\end{equation}
where $\tilde g$ denotes the coupling coefficient between the qubit of the system and the qubit of the added particle.
Above equation shows that the coherence can be preserved by using a single extra qubit and a single extra field (i.e.
the  particle-pairing approach \cite{duan97}) if the Rabi frequency of the driving field satisfies the above condition
Eq. (\ref{eq1:Memory-driving}).

As a second example one may also store a qubit by employing a sequence of pulses to drive the system as
demonstrated in \cite{viola97}.
In this case, the Hamiltonian $H_{fs}$ in the Schr\"odinger picture is expressed as
\begin{equation}\label{eq1:Memory-HFS}
H_{fs}=\sum^{n_p}_{n=1}V^{(n)}(t)e^{-i\omega(t-t_p^{(n)})}\sigma^++H.c.,
\end{equation}
where $n_p$ is the number of pulses, $t_p^{(n)}=(n-1)(\tau_p+\Delta t)$ for
$n\in \{1,2,\cdots, n_p\}$ and
\begin{equation}\label{eq1:Memory-sequence}
V^{(n)}(t)=\left\{
\begin{array}{rr}
V& \,\,\,\,\,\,\, t^{(n)}_p\leq t\leq t^{(n)}_p+\tau_p\\ %
0  &\,\,\,\,\,\, elsewhere,
\end{array}
\right.
\end{equation}
which means that a pulse is applied at each instant $t=t_p^{(n)}$ with pulse
duration $\tau_p$ and separation $\Delta t$ between pulses. Transferring
Eq.(\ref{eq1:Memory-HFS}) into the interaction picture gives
\begin{eqnarray}\label{eq2:Memory-HFS}
H_{fs}&=&\sum^{n_p}_{n=1}V^{(n)}(t) e^{i\omega(\sigma_z/2) t_p^{(n)}}\sigma_z e^{-i\omega(\sigma_z/2)t_p^{(n)}}
\nonumber\\ &=& \sum^{n_p}_{n=1}V^{(n)}(t) e^{i\omega t_p^{(n)}}\sigma^++H.c..
\end{eqnarray}

Then the Rabi frequency has the following form
\begin{equation}\label{eq3:Memory-Rabi}
\Omega=V\sum^{n_p}_{n=1}e^{i\omega (n-1)(\tau_p+\Delta t)}.
\end{equation}
Since $K=0$ and $L=1$ in this case, Eq. (\ref{eq1:Memory-condition}) and the above equation give
\begin{equation}\label{eq2:Memory-driving}
V\left|\sum^{n_p}_{n=1}e^{i\omega (n-1)(\tau_p+\Delta t)}\right|=\sqrt{\frac{\hbar^2 D\gamma}{4}},
\end{equation}
which means that for driving field amplitudes satisfying Eq. (\ref{eq2:Memory-driving}) a DFS, i.e. coherence
preservation, can be guaranteed. When $n_p\rightarrow \infty$, we obtain the simplified expression
\begin{equation}\label{eq3:Memory-driving}
V=\sqrt{\hbar^2D\gamma \sin^2\frac{\omega}{2}(\tau_p+\Delta t)}.
\end{equation}
We note that in the particular case $\tau_p+\Delta t=n2\pi/\omega=nT (n=0,1,\cdots, )$ the pulse amplitude satisfies
$V=0$. However, since the high order terms in the master equation (\ref{eq1:DFS-master}), which are very small
contrast to the linear term, have been omitted, accordingly one should have actually $V\approx 0$. Thus coherence
preservation can be guaranteed by a very small pulse amplitude in this situation.

Based on Eq. (\ref{eq1:Memory-condition}) we shall now present a third example which has not been explored in the
literature before. We consider the situation  $\tilde g_{sk}=\tilde g_0 (k\in\{1,2,\cdots, K\}$ and $L=0$, which means
that the information qubit is surrounded by $K$  additional qubits with identical coupling strengths. We thus may
imagine the information qubit to be in the center of a sphere being surrounded by the $K$ additional qubits situated
on the surface of this sphere. The sphere will certainly modify the effective Hamiltonian of the information system and
subsequently turn out to cancel the passive effects from the bath $\mathcal B$. Alternatively we may regard this
sphere to form an auxiliary part of the environment which engineers the bath $\mathcal B$. At an appropriate
condition, the coherence of the information system may be preserved. This condition can be obtained from Eq.
(\ref{eq1:Memory-condition}). With $L=0$ and the coupling coefficients between the information qubit and the
additional qubits being $g_0$, we obtain
\begin{equation}\label{eq2:Memory-condition}
2K\Im (\tilde g_0)=\hbar D.
\end{equation}
In general $D$ is a constant, so that above equation represents a relationship between the number of the additional
qubits and the coupling strength of the additional qubits and the information qubit.
Eq. (\ref{eq2:Memory-condition}) also shows that a driving field is not necessary to inhibit decoherence in quantum
memories.

\subsection{Decoherence-free quantum operations}

Up to now we have put forward means to refrain a system from evolving at all and this in spite of a
surrounding bath. Now we do want a system to evolve, however, under a well defined precise operation.
The inhibition of decoherence during a quantum gate operation is governed again by Eq. (\ref{eq2:MOD-H'})
however only if $\epsilon=1$. Then the additional Hamiltonian $\mathcal H_p$ is associated with the
Hamiltonian $\mathcal H_g$, decribing the gate operation.
We now pursue by deriving a practical condition for inhibiting decoherence during a gate operation with
the help of extra qubits and extra fields.

We consider a single qubit quantum gate, which can be described by a 2-dimensional
unitary matrix. In general, an arbitrary single qubit quantum gate can be
expressed as \cite{nielsen00},
\begin{equation}\label{eq1:Gate-gate}
\mathcal U=\left(
\begin{array}{cc}
e^{i(\alpha-\beta/2-\delta/2)}\cos\frac{\vartheta}{2} &
-e^{i(\alpha-\beta/2+\delta/2)}\sin\frac{\vartheta}{2}  \\ %
e^{i(\alpha+\beta/2-\delta/2)}\sin\frac{\vartheta}{2} &
e^{i(\alpha+\beta/2+\delta/2)}\cos\frac{\vartheta}{2}
\end{array}
\right ).
\end{equation}
where the real numbers $\alpha, \beta, \delta$ and $\vartheta$ are the characteristic parameters of
the quantum logic gate. Then $H_g$ can be expressed as
\begin{equation}\label{eq1:Gate-Hg}
H_g=\frac{\mathcal U-\cos(\frac{\tau}{\hbar}) \mathcal I}{i\sin(\frac{\tau}{\hbar})},
\end{equation}
where $\mathcal I$ is the 2-dimensional unity matrix and $\tau$ is the time during the action of the quantum gate on
the qubit. Then Eq. (\ref{eq3:MOD-H'}) gives,
\begin{equation}\label{eq2:Gate-H11}
H_{11}=\sum^K_{k=1}\tilde g_{sk}+\frac{e^{i(\alpha-\beta/2-\delta/2)}
\cos\frac{\vartheta}{2}}{i\sin(\frac{\tau}{\hbar})} - \frac{\cos(\frac{\tau}{\hbar})}{i\sin(\frac{\tau}{\hbar})},
\end{equation}
\begin{equation}\label{eq2:Gate-H22}
H_{22}=\sum^K_{k=1}\tilde g_{sk}+\frac{e^{-i(\alpha-\beta/2-\delta/2)}
\cos\frac{\vartheta}{2}}{i\sin(\frac{\tau}{\hbar})} - \frac{\cos(\frac{\tau}{\hbar})}{i\sin(\frac{\tau}{\hbar})},
\end{equation}
\begin{equation}\label{eq2:Gate-H12}
H_{12}=\sum^L_{l=1}\tilde\Omega_l-e^{i(\alpha-\beta/2+\delta/2)}\sin\frac{\vartheta}{2},
\end{equation}
and
\begin{equation}\label{eq2:Gate-H21}
H_{21}=H^*_{12}.
\end{equation}
Eqs. (\ref{eq2:Gate-H11}) and (\ref{eq2:Gate-H22}) lead to
\begin{equation}\label{eq2:Gate-DH}
\Delta H=2i\sum_{k=1}^K\Im(g_{sk})+\nu, \,\,\,\, \kappa=\frac{4}{\hbar}\sum_{k=1}^K\Im(g_{sk}),
\end{equation}
where $\nu={2\cos\frac{\vartheta}{2}\sin\Theta}/{\sin(\frac{\Delta t}{\hbar})}$ and $\Theta=\alpha-\beta/2-\delta/2$.
Using the same treatment as in the previous subsection for quantum memories, we finally obtain
\begin{equation}\label{eq1:gate-condition}
\left|\sum^L_{l}\tilde\Omega_l\right|=\cos\Theta'\sin\frac{\vartheta}{2}+
\sqrt{\frac{\gamma(\frac{\hbar^2}{4}\mu^2+\nu^2)}{2\mu}},
\end{equation}
where $\Theta'=\alpha-\beta/2+\delta/2$. Obviously, the quantum gate will
influence the choice of the additional system $\cal Q$ and $\cal F$ aimed to
inhibit the decoherence.

As an example, we consider the case of the quantum gate being the Hadamard gate
\begin{equation}
\mathcal H=\frac{1}{\sqrt{2}}\left(
\begin{array}{cc}
1&1\\ 1&-1
\end{array}
\right).
\end{equation}
In this situation, one can easily calculate the characteristic parameters of the quantum gate through Eq.
(\ref{eq1:Gate-gate}), i.e., $\alpha=\pi/2, \beta=0, \delta=\pi$ and $\vartheta=\pi/2$, subsequently $\Theta=0,
\Theta'=\pi$. Then the parameters of the additional systems $\cal Q$ and $\cal F$ obey the following relationship
\begin{equation}\label{eq2:gate-condition}
\left|\sum^L_{l}\tilde\Omega_l\right|=\sqrt{\frac{\hbar^2\gamma\mu}{8}}-\frac{\sqrt{2}}{2}.
\end{equation}
Comparing the above equation with Eq. (\ref{eq1:Memory-condition}), one can find easily that the difference is only an
additional constant. In the case $K=1, L=1$, i.e., the particle-pairing approach, we have
\begin{equation}\label{eq3:gate-condition}
|\tilde\Omega|=\sqrt{\frac{\hbar\gamma(\hbar D-2\Im(\tilde g))}{4}}-\frac{\sqrt{2}}{2}.
\end{equation}

In the case if the system is driven by a sequence of pusles as described in Eq. (\ref{eq3:Memory-Rabi}),
one has the following condition
\begin{equation}\label{eq4:gate-condition}
V\left|\sum^{n_p}_{n=1}e^{i\omega(n-1)(\tau_p+\Delta t)}\right|=\frac{\hbar}{2}\sqrt{D\gamma}-\frac{\sqrt{2}}{2},
\end{equation}
where the parameters are the same as in Eq. (\ref{eq1:Memory-sequence}).

Finally we consider the model proposed in the last paragraph of Sect. \ref{mem}. Since in this situation $\tilde
g_{sk}=\tilde g_0$ and $L=0$, one can easily obtain the following equation
\begin{equation}\label{eq5:gate-condition}
2K\Im (\tilde g_0)=\hbar D-\frac{2}{\hbar \gamma}.
\end{equation}
Comparing this condition to Eq.(\ref{eq2:Memory-condition}) we find that there is a modification term
which is induced by the quantum operation. Thus also a quantum operation may be decoherence-free solely due
to placing an ensemble of qubits around the information qubit to be prevented from decoherence.

\section{Conclusions}
\label{con}

The phenomena of coherence and decoherence have been investigated in detail for a two-state quantum system (information
system) in a thermal bath modeled by an infinite-mode harmonic oscillator.
In Sect. \ref{dyn} we confirmed the general notion that decoherence arises from the coupling between the information system
and the environment, i.e., the entanglement between the single mode of the system and the infinite number of modes of the
bath. The entanglement, if not compensated, leads the irreversible loss of coherence.

In order to prevent the loss of coherence of the system an assistant system (or process) was introduced in Sect.
\ref{workeq} as a means of generating a DFS for the information system. We derived general conditions which the
corresponding extra Hamilton operator has to fulfill in order to decouple the information system from the environment
and thus to inhibit decoherence. The assistant Hamilton operator was assumed general up to the only restriction of
inducing a weak coupling to the system and at this stage it may be associated with either physical or unphysical
situations.

Finally in Sect. \ref{app} we investigate possible applications of the obtained condition in quantum information
processing. In this case we restricted ourselves to assistant Hamilton operators being described by an additional set
of qubits and an additional set of classical fields.
In the two important examples of quantum information storing and quantum gates, our general condition showed various
situations where those processes can be prevented from decoherence by rather practical means. Precise conditions
on the free external system parameters are derived involving the situations where extra qubits or extra fields are
employed alone and where combinations are used.
This way previous results on decoherence-free quantum information processing were confirmed and a previously unexplored
mechanism was put forward. This involves the placement of the information qubit into the center of a sphere with
additional identical qubits on its surface. The coupling constants of the extra qubits can be adapted to the
interaction of the bath on the information qubit such that its decoherence is inhibited.

\vspace{0.2cm}
\section*{Acknowledgements}

We have benefited from useful discussions with J\"org Evers and Michael Fleischhauer. This work is supported by the
Alexander von Humboldt-Stiftung foundation (Grant Nr. IV CHN 1069575 STP) and the German Science Foundation
(Nachwuchsgruppe within SFB276).

\end{document}